\documentclass{elsart}
\begin{document}
\input{psfig}
\def\Flam{F_{\lambda_0}}
\def\dlam{d\lambda}
\def\jlam{j(\lambda_e,z)}
\def\lame{\lambda_{\rm e}}
\def\Egam{E_{\gamma}}
\def\eebl{\epsilon_{EBL}}
\def\gsim {\lower .1ex\hbox{\rlap{\raise .6ex\hbox{\hskip .3ex
        {\ifmmode{\scriptscriptstyle >}\else
                {$\scriptscriptstyle >$}\fi}}}
        \kern -.4ex{\ifmmode{\scriptscriptstyle \sim}\else
                {$\scriptscriptstyle\sim$}\fi}}}
\def\lsim {\lower .1ex\hbox{\rlap{\raise .6ex\hbox{\hskip .3ex
        {\ifmmode{\scriptscriptstyle <}\else
                {$\scriptscriptstyle <$}\fi}}}
        \kern -.4ex{\ifmmode{\scriptscriptstyle \sim}\else
                {$\scriptscriptstyle\sim$}\fi}}}

\newcommand{\beq}{\begin{equation}} 
\newcommand{\eeq}{\end{equation}} 
\newcommand{\beqa}{\begin{eqnarray}} 
\newcommand{\eeqa}{\end{eqnarray}} 
\newcommand{\micron}{\mu {\rm m}} 
\runauthor{J.S. Bullock, R.S. Somerville, D. MacMinn, and J.R. Primack}
\begin{frontmatter}
\title{Constraining the IMF using TeV gamma ray absorption}
\author[UCSC]{J.S. Bullock}
\author[J]{R.S. Somerville}
\author[C]{D. MacMinn}
\author[UCSC]{and J.R. Primack}

\address[UCSC]{Physics Department, University of California, Santa Cruz}
\address[J]{Racah Institute of Physics, Hebrew University, Jerusalem}
\address[C]{Deceased}
\begin{abstract}

Gamma rays from distant sources suffer attenuation due
to pair production off of $\sim 1 \micron$ EBL photons via 
$\gamma \gamma \rightarrow e^{+}e^{-}$~\cite{gould}.  
The cross section for the pair production
is maximized just above threshold, when 
$\lambda_{EBL} \sim  (E_{\gamma}/TeV) \micron$.  We may exploit
this process in order to indirectly measure the EBL,
and constrain models of galaxy formation~\cite{SDS,MP,BSMP,SBP}.
Here, using semi-analytic models of galaxy formation, we
examine how gamma ray absorption may be
used as an indirect probe of the stellar initial mass
function (IMF), although
there is a degeneracy with dust modeling.  We point
out that with the new generation of gamma ray telescopes
including STACEE, MAGIC, HESS, VERITAS, and
Milagro, we should soon possess a wealth of new data
and a new method for probing the nature of the IMF.

\end{abstract}
	

\end{frontmatter}

\begin{figure}[h] 
\centerline{\psfig{file=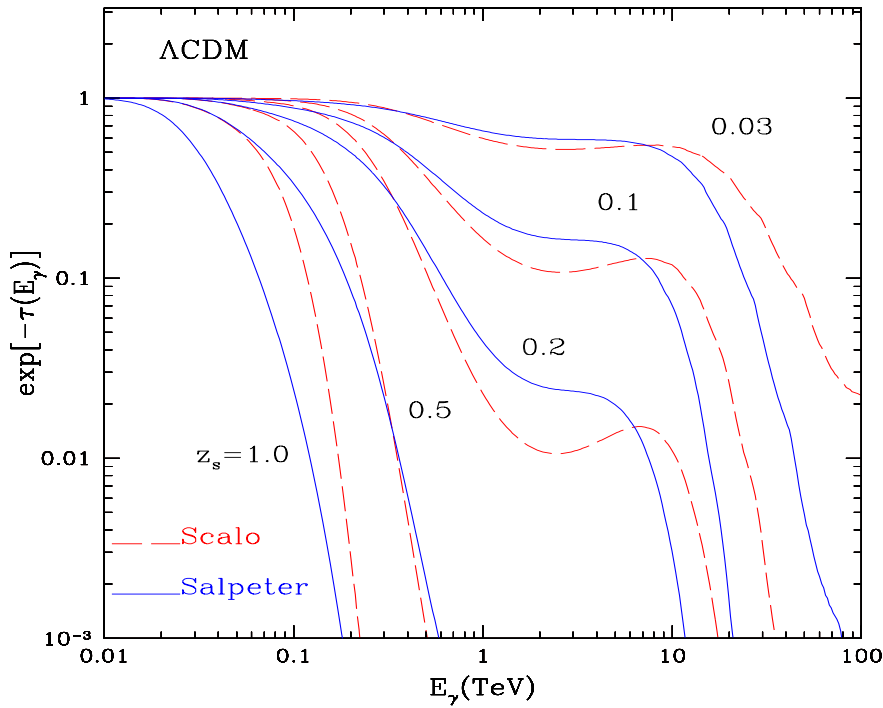,width=12truecm,height=10truecm}}
        \caption{\label{figure} The attenuation factor for 
$\Lambda$CDM models using a Salpeter (solid) and Scalo (dashed) IMF
as a function of gamma ray energy $E_{\gamma}$, where  $\tau(E_{\gamma})$
is the optical depth of the universe.
The Salpeter IMF produces more high mass stars, thus more 
ultraviolet light and more $\sim 100\mu$m light reradiated
by dust. For distant sources (redshift $z_s \sim 1$) the
Salpeter model's increased UV light causes a noticeably 
larger attenuation of gamma rays.
For nearby sources, the increase in reradiated light relative to Scalo
implies significantly increased gamma ray 
attenuation at $E_{\gamma} \gsim 10$TeV.
}

\end{figure}

\section{Calculating the EBL and constraining the IMF}

Semi-analytic merging-tree (SAM) models of galaxy formation
incorporate parameterized 
treatments of astrophysical processes such as 
gas cooling, star formation, supernovae feedback and dust absorption within the
hierarchical structure formation scenario.  Using the SAM models developed in
Ref.~\cite{SP}, we can efficiently model the origin of the EBL in a variety of
cosmological scenarios in a physical way. We can then use observations to
contrain the nature of the IMF and the effects of dust (see ~\cite{BSMP,SBP} 
and ~\cite{PBSM}, hereafter PBSM, for
details).

In this contribution we focus on 
a model set in a 
flat $\Lambda$CDM 
universe ($\Omega_m = 0.4$, $\Omega_{\Lambda}=0.6$),
and investigate how the nature of the IMF affects the
expected gamma-ray attenuation.
We present results using two commonly used forms for the IMF, Scalo
\cite{scalo:86} and Salpeter \cite{salpeter:55}. Our results will primarily
constrain the ratio of high mass to low mass stars produced. This quantity is
of general interest in the context of understanding supernovae rates, metal
production, and high redshift galaxies which are typically identified in the
far UV. Note that there is some degeneracy between the effects of the IMF and
the wavelength dependence of the dust extinction curve (see PBSM for a
description of our treatment of dust).

Figure 1 shows the gamma ray attenuation factor, $\exp(-\tau)$, from our
$\Lambda$CDM models using a Salpeter (solid) and Scalo (dashed) IMF.
The optical depth 
is a function of gamma ray energy, $\tau(E_{\gamma})$, and is most strongly
influenced by the EBL at wavelengths $\lambda_{EBL} \sim (E_{\gamma}/{\rm
TeV})\mu{\rm m}$, where the cross section for pair production is maximized.
The numbers next to each pair of curves 
indicate the redshift of the source, $z_s$.

As discussed in detail in PBSM (see figures 2 and 4), the Salpeter IMF produces
more high mass stars, thus more ultraviolet light than does the Scalo IMF, and
therefore a larger optical depth to $\lsim 1$TeV gamma rays.  The increased UV
light, in addition, produces more $\sim 100\mu$m light reradiated by dust, and
a larger optical depth to $\gsim 10$TeV photons.  Distant sources ($z_s \gsim
0.5$) should provide an interesting probe of the IMF using $\lsim1$TeV gamma
ray telescopes.  For nearby sources, the excess reradiated light relative to
Scalo implies excess gamma ray attenuation at $E_{\gamma} \gsim 10$TeV, the
relevant range for ground-based air-shower detectors.

\section{Conclusions}

Gamma ray attenuation at $E_{\gamma} \lsim 1$TeV and $\gsim 10$TeV is
significantly affected by the IMF, 
specifically the ratio of high to low mass
stars, although there 
is some degeneracy
associated with uncertainties in 
the modeling of dust absorption and reradiation.  Observations of gamma ray
absorption below $\sim 1$TeV for high redshift sources ($z_s \gsim 0.5$) and
above $\sim 10$TeV for nearby sources ($z_s \sim 0.03$) will provide a useful
probe of the nature of the IMF and galaxy formation.  For more details,
including fitting functions of the optical depth of the universe as a function
of redshift, cosmology, IMF, and gamma ray energy, see~\cite{BSMP}.

\end{document}